\documentclass{appolb}
\usepackage{graphicx,amsmath,amssymb}

\usepackage[mathscr]{euscript}

\renewcommand{\vec}{\boldsymbol}
\newcommand{\beq}{\begin{equation}}
\newcommand{\eeq}{\end{equation}}
\newcommand{\bea}{\begin{eqnarray}}
\newcommand{\eea}{\end{eqnarray}}
\newcommand{\baa}{\begin{array}}
\newcommand{\eaa}{\end{array}}

\def\eq#1{{Eq.~(\ref{#1})}}

\def\fig#1{{Fig.~\ref{#1}}}
\newcommand{\intl}{\int\limits}

\newcommand{\bas}{\bar{\alpha}_S}

\newcommand{\nn}{\nonumber}

\newcommand{\h}{\frac{1}{2}}

\newcommand{\x}{\vec{x}}
\newcommand{\vb}{\vec{b}}
\newcommand{\z}{\vec{z}}

\newcommand{\Lb}{\left(}
\newcommand{\Rb}{\right)}

\newcommand{\pp}{\partial}

\renewcommand{\vec}[1]{\boldsymbol{#1}}

\newcommand{\dY}{\delta \tilde{Y}}

\begin{document}
\title{Modified Homotopic approach for diffractive production%
\thanks{Presented at ``Diffraction and Low-$x$ 2024'', Trabia (Palermo, Italy), September 8-14, 2024.}  
}
\author{Carlos Contreras, Jos\'e Garrido \\
\address{Departamento de F\'isica, Universidad T\'ecnica Federico Santa Mar\'ia,  Avda. Espa\~na 1680, Casilla 110-V, Valpara\'iso, Chile}
\\[3mm]
{Eugene  Levin \\
\address{Department of Particle Physics, School of Physics and Astronomy,
Raymond and Beverly Sackler
 Faculty of Exact Science, Tel Aviv University,  Israel}
 }
\\[3mm]
 Rodrigo Meneses
\address{Escuela de Ingenier\'ia Civil, Facultad de Ingenier\'ia, Universidad de Valpara\'iso, Avda  General Cruz 222 , Valpara\'iso, Chile}
}
\maketitle
\begin{abstract}
We review the recent developments of the use of the homotopy method for solving the non-linear evolution equation for the diffractive production in deep inelastic scattering. We introduce   part of the non-linear corrections in the linear term. This simplified non-linear evolution equation is solved analytically taking into account the initial and boundary conditions for the process.   It turns out that these corrections are rather small and can be estimated in the regular iterative procedure.
\end{abstract}
  
\section{Introduction}
In this paper we   present  a  procedure to  solve the non-linear equations for  diffractive process that in  QCD
govern the dynamics in the saturation region. 
In our previous paper~\cite{CLMNEW, CLMNEW2} we found a solution of the Balitsky-Kovchegov (BK) equation~\cite{BK1, BK2} that gives the dipole scattering amplitude using the homotopy approach. It has been shown that this approach allows us to collect all essential contribution into the linear equation which can be solved analytically,  and to  propose an iteration  procedure, which is partly numerical and leads to small corrections. 

For diffractive production we have   several different kinematic regions for $N^D$, where 
$
 \sigma_{\rm dipole}^{\rm diff}(r_{\perp},Y, Y_0)
\,=\int\,d^2 b\,N^D(r_{\perp},Y, Y_0;\vec{b})\,
$
is the cross section of diffractive production with the rapidity gap larger than $Y_0$.
The non-linear evolution equation for $N^D\Lb Y, Y_0,r_{10}; b \Rb$ that describes these    diffraction production in deep inelastic scattering  has been derived in Ref.~\cite{KOLEB} and has the following form:
 \bea \label{SDEQ}
&&\frac{\partial N^D( Y, Y_0,r_{10}; b)}{\partial Y}=
\frac{\bas}{2 \pi}\int d^2 r_2 \frac{ x^2_{01}}{x^2_{02}\,x^2_{12}} \{N^D_{12}  
 \nn  \\
&+& N^D_{20}  - N^D_{10}  + N^D_{12}  N^D_{20} 
-\,2 N^D_{12} N_{20}     - 2 N_{12} N^D _{20}   +\,2 N_{12}N_{20} \},
 \eea    
where $N_{i k}=N(Y;r_{i k}; b) $ is the elastic scattering amplitude of a dipole with size $r_{i k}$ and rapidity $Y$ and  $N^D_{i k}=N^D(Y; Y_0, r_{i k}; b) $ is the cross section of the diffractive production with the rapidity gap larger than $Y_0$ at the impact parameter $b$ for the same scattering process. 

 \begin{figure}
 	\begin{center}
 	\leavevmode
 		\includegraphics[width=6cm]{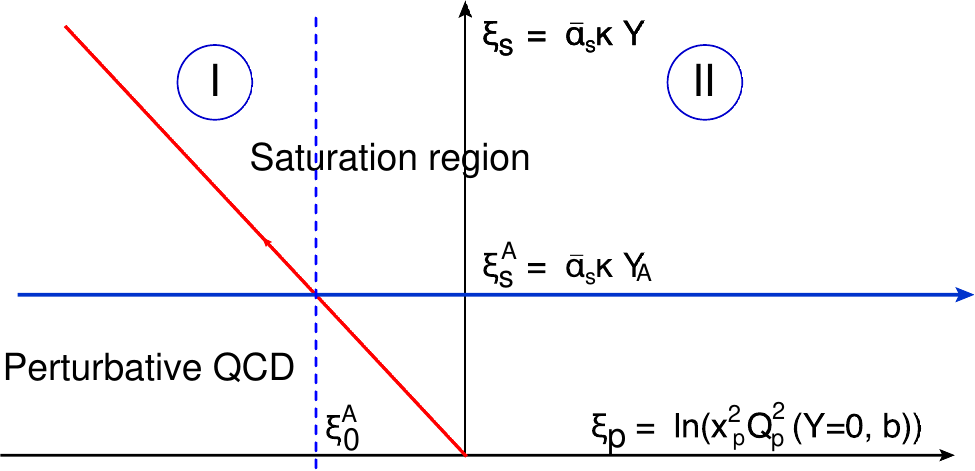}
 	\end{center}
 	\caption{Saturation region of QCD for the elastic scattering amplitude with   $\kappa= 
  \frac{\chi\Lb \gamma_{cr}\Rb}{1 - \gamma_{cr}} $ and  $\chi\Lb \gamma\Rb$  is the BFKL characteristic function. The critical line ($z=0$) is shown in red. The initial condition for scattering   is given at $\xi_s = 0$ and for heavy nuclei the initial condition  is shown by the  blue line. At the vertical blue  dotted line $\xi = \xi^A_0$ the amplitudes described in region I and II are matched.} 
 	\label{sat}
 \end{figure}
 
 In general, the homotopy method can be used   as an  effective procedure for solving an equation
 of the form $
\mathscr{L}[u] +  \mathscr{N_{L}}[u]=0
$,  where the linear part $\mathscr{L}[u] $ is a differential  or integral-differential operator, and the  non-linear part $
 \mathscr{N_{L}}[u] $ has an arbitrary form. To solve it, we introduce  the following  equation for the homotopy function $ {\mathscr H}\Lb p,u\Rb$:
${\mathscr H}\Lb p,u\Rb\,\,=\,\,\mathscr{L}[u_p] \,+ \,  p\, \mathscr{N_{L}}[u_p] \,\,=\,\,0
 $.   Solving this expression, we reconstruct the function
 \beq \label{HOM3}
 u_p\Lb Y,  \x_{10},  \vb\Rb= u_0\Lb Y,  \x_{10},  \vb\Rb+p\, u_1\Lb Y,  \x_{10},  \vb\Rb \,+\,p^2\, u_2\Lb Y,  \x_{10},  \vb\Rb \,\,+\,\,\dots
 \eeq
 with $\mathscr{L}[u_0] = 0$. \eq{HOM3}  gives  the solution to the non-linear equation for $p = 1$.  The hope is that several  terms in this series    will give a good  approximation to the solution of the non-linear  equation. 
  
 The linear equation is obvious in the perturbative QCD region (see \fig{sat}), where it is the BFKL equation. However,  we will show that inside the saturation region (see \fig{sat})  we can find the linearized  equation based on the approach of Ref.~\cite{LETU}.  We  demonstrate that the non-linear term, which includes the remains of the non-linear corrections, leads to small contributions and can be treated in the perturbation approach.  This aspect  has been  discussed in Ref.~\cite{CLMNEW2}, whose results are here briefly described.
 
 The kinematic region where we are looking for the solution   for the scattering amplitude of a dipole ($x_{10} =r $)   with  a nucleus target in the plot with $\xi_s$ and $\xi$ axes is shown in \fig{sat}, where
 $\xi=\ln\Lb x^2_{10}\,Q^2_s( Y=Y_A, b )\Rb$ and $ \xi_s = \ln(Q^2_s(Y)/Q^2_s( Y=Y_A, b))$.
 One can see that for $z\, <\,0$  we have
 the perturbative QCD region where the non-linear corrections are small and we can safely use the BFKL linear equation for the scattering amplitude. The geometrical scaling variable  $z$ is defined as
 $z= \ln( r^2 \,Q^2_s( \dY, b))=\xi_s+\xi $,
  with  the saturation moment  $Q^2_s(Y) =Q^2_s\Lb Y_A\Rb e^{\bas\,\kappa \Lb Y - Y_A\Rb}$, $\kappa=4.88$,  and $Y_A$ denotes $ Y_A = \ln A^{1/3}$, where $A$ is the number of nucleons in a nucleus.

For $z>0$ the non-linear corrections become essential and we enter the saturation region. For the scattering with nuclei,  the saturation region can be divided in two parts. For $\xi <\,\xi^A_0$ the amplitude has the geometric scaling behaviour~\cite{CLMNEW} and depends only on one variable, $z$. For $\xi\, >\,\xi^A_0$ 
this geometric scaling behaviour is broken. 
This process can be characterized by two kinematic regions: for $ r_{\perp} Q_s( Y_0) < 1 $  and $ r_{\perp} \,Q_s\Lb Y - Y_0 \Rb\,<\,1$ we can replace $N$  by the BFKL Pomeron.
For $ r_{\perp} \,Q_s\Lb Y_0 \Rb\,>\,1$  and $ r_{\perp} \,Q_s\Lb Y - Y_0 \Rb\,<\,1$  the elastic amplitude is in the saturation region and the production of gluons can be computed using the BFKL Pomeron exchange.
 Finally, $ r_{\perp} \,Q_s\Lb Y_0\Rb \,>\,1$  and $ r_{\perp} \,Q_s\Lb Y - Y_0\Rb \,>\,1$ 
 is the kinematic region where non-linear corrections  for gluon production are essential.


 It turns out that \eq{SDEQ} can be rewritten in a simple form 
 introducing a new function,
\begin{equation}\label{AMPLITUD}
	\mathscr{ N}(z,\dY, \delta Y_0)\, =\,  2\, N(z,\dY)\, -\, N^D(z, \dY, \delta Y_0)\;,
\end{equation}
where  $N\Lb z, \dY\Rb$ is the solution of the BK equation. The new variables $\dY$ and $\delta Y_0$ are defined as $\dY= \bas\Lb Y - Y_A\Rb$ and $\delta Y_0 = \bas \Lb Y_0 - Y_A\Rb$.
This function has clear physics meaning: the inelastic cross section of all events with rapidity gap from $Y=0$ to   $Y=Y_0$. Then,    \eq{SDEQ} in terms of the  function ${\mathscr N}$ takes the form of the BK equation, viz.
 \beq \label{SDBK}
 \frac{\partial{\mathscr N}_{01}}{\partial Y}\,=\,\bas\int \frac{d^2\,x_{02}}{2 \pi} \frac{ x^2_{01}}{x^2_{02}\,x^2_{12}}\Big\{ {\mathscr N}_{02} + {\mathscr N}_{12} - {\mathscr N}_{02} {\mathscr N}_{12} - {\mathscr N}_{01}\Big\},
 \eeq 
with the  initial conditions
 $ {\mathscr N}(z \to z_0,\dY = \delta Y_0  , \delta Y_0)  =  2 N(z_0,\delta Y_0) - N^2(z_0,\delta Y_0)
$; it  holds only in the region I, while in region II we have to use a more general expressions for the elastic scattering amplitudes (see Ref.~\cite{CLMNEW}).

Our strategy for finding solution looks as follows: first we   solve \eq{SDBK} with $\mathscr{ N}_{01} \,=\, 1 - \Delta^D$, and after that we return to \eq{SDEQ}.  
\section{Modified homotopy approach for $\Delta^D_0$ and numerical estimation for  $\Delta_1^D$  }  
\begin{figure}
 	\begin{center}
	\begin{tabular}{c c} 
 	\leavevmode
 		\includegraphics[width=6cm]{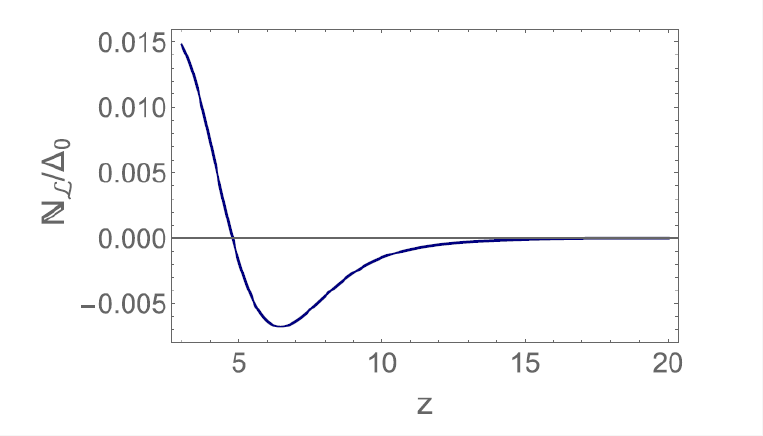}& \includegraphics[width=6cm]{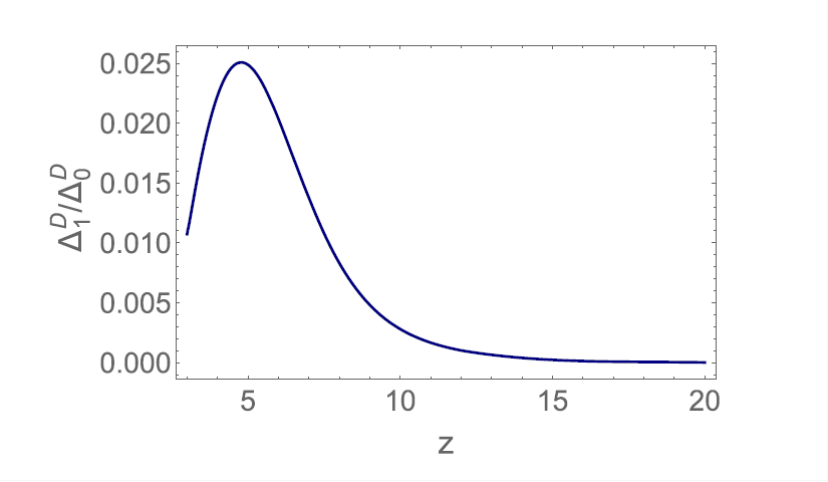} \\
	\fig{m}-a& \fig{m}-b\\
	\end{tabular}	
	\end{center}
 	\caption{\fig{m}-a describes the ratio $ \mathscr{N_{L}}[\Delta^D_{0}]/\Delta^D_0$.  \fig{m}-b describes
          $\Delta^D_{1}/\Delta^D_0$   versus $z$ for $z > z_0$. $z_0$ is taken to be 3, $\bas = 0.2$. }
 	\label{m}
 \end{figure}

  Including  part of the non-linear term into the definition of $ \mathscr{L}[u_0] $, we can  find the solution to non-linear equation of \eq{SDBK}, suggested in Refs.~\cite{ CLMNEW}.      
      Following the main ideas of Ref.~\cite{LETU}, we solve \eq{SDBK} replacing $ \mathscr{N}\Lb z, \delta Y_0\Rb $ by  $ 1-\Delta^D\Lb  z, \delta Y_0\Rb$. For  this function the equation takes the form
      \beq\label{SDBK1}
 \frac{\partial  \Delta^D_{01}}{\partial Y}\,=\,\bas\int \frac{d^2\,x_{02}}{2 \pi} \frac{ x^2_{01}}{x^2_{02}\,x^2_{12}}\Big\{  \Delta^D_{02} \Delta^D_{12} - \Delta^D_{01}\Big\}\;,
\eeq 
with the corresponding  initial conditions for $\Delta^D_{01}$.   
   We suggest  to simplify the non-linear term replacing  it by
    \beq \label{SDBK12}
\int \frac{d^2\,x_{02}}{2 \pi} \frac{ x^2_{01}}{x^2_{02}\,x^2_{12}} \Delta^D_{02} \Delta^D_{12} \,\to\,\Delta^D_{01} \intl^z_0 d z'  \Delta^D_{02}\,\,=\,\,  \Delta^D_{01} \Bigg( \zeta - \intl^\infty_z d z'  \Delta^D_{02}\Bigg)\;.\eeq    
 This contribution stems from the region $x_{02}\,\ll\,x_{01}$ (see Ref.~\cite{LETU}) and, to find the solution $\Delta^D_0$, we   need to solve  the equation $  \mathscr{L}\Lb \Delta^D_0\Rb\,\,=\,\,0$.
Introducing  $\Delta_0^{D}\Lb z,  \xi_s\Rb   = \exp\Lb - \Omega^{(0)}\Lb z, \xi_s\Rb\Rb$, we can obtain    the next general equation ($\xi_s = \kappa \,\dY$),
\beq\label{GSS2}
\kappa \dfrac{\pp^2\Omega^{(0)}}{\pp \xi_s \,\pp z} =1 -  e^{ - \Omega^{(0)}\Lb \z,\xi_s\Rb}\;,
\eeq
which in the region I  has  traveling wave solution (Ref.~\cite{MATH2}) with  the geometric scaling  behaviour, that
can be found from the following implicit  equation:
\beq\label{GSS5}
\mathscr{U}\Lb\Omega^{(0,I)}, \Omega^{(0)}_0 = a\Rb =	\intl^{\Omega^{(0)}}_{\Omega^{(0)}_0}\frac{ d \Omega'}{\sqrt{ \Omega' + \exp\Lb - \Omega'\Rb -  \Omega^{(0)}_0}}=\sqrt{\frac{2}{\kappa}}\,\Lb z + C_2\Rb,
	\eeq
$C_2$  as well as $\Omega^{(0)}_0$   can be found using the boundary conditions. We can 
solve the  \eq{GSS5}  for  $\Omega^{(0,I)}\Lb z \Rb $, and  obtain  
  $\Delta^{(0,I)}_0$ which is given by
    \beq \label{GSS14}  
\Delta^{(0,I)}_0\Lb z\Rb  =\Delta_{LT}\Lb z\Rb\exp \Lb - a  + \sum^\infty_{k=1} \frac{(-1)^{k-1}(2k -1)!! (2 \kappa)^{k +\h}}{2^k k\,k!  \Lb z - \tilde{z}\Rb^{2(k +\h)}}   \Delta^k_{LT}\Lb z \Rb \Rb \;,
    \eeq 
     where $ \Delta_{LT}\Lb z\Rb    \,=\,\exp( - \frac{\Lb z \,-\,\tilde{z}\Rb^2}{2\,\kappa})$ is the Levin-Tuchin solution~\cite{LETU}.  
  In the region II,  $ \Omega^{(0,II)}\Lb z,\xi_s \Rb$  is described by the equation
 \beq \label{GSG1}
\dfrac{\pp^2\Omega^{(0,II)}\Lb z, t\Rb}{\pp z^2} \,\,-\,\,\dfrac{\pp^2\Omega^{(0,II)}\Lb z,t\Rb}{\pp t^2}	\,\,=\,\,\frac{1}{\kappa}\Lb 1 -  e^{ - \Omega^{(0,II)}\Lb z, t\Rb}\Rb \;,
\eeq
where $z = \xi_s + \xi$ and $t = \xi_s - \xi$, whose solution is given again  by the implicit  traveling wave solution  $\mathscr{U}\Lb\Omega^{(0,II)},  \Omega_0\Rb $
   (see  \eq{GSS5}). Therefore  for    $ \Delta^{(0,II)}( z, t) $ we have  
   \beq \label{GSG7}  
 \Delta^{(0,II)}=\Delta_{LT}\exp\Lb -a  +  \sum^\infty_{k=1} \frac{(-1)^{k-1}(2 k -1)!!\,(2 \kappa)^{k +\h}}{2^k k\,k! \,( (1+\nu) z + \nu t - \hat{z})^{2(k +\h)}} \Delta^k_{LT}\Rb. 
    \eeq 
Now,    we are in condition to use the homotopic approach  to obtain the first correction $\Delta^D_1$ from the general equation valid in the region I and II:
    \beq \label{SDBK2}
                 \Lb \kappa \frac{\pp}{\pp z} +\,\,z\, \Rb\,\Delta^D_1\Lb z, \xi, z_0\Rb \,=\,-    \mathscr{N_{L}}[\Delta_0^D]\;, 
  \eeq
where the non-linear contribution is given by
 \beq \label{2I1}
\mathscr{N_{L}}[\Delta^D]\,\,=\,\,\bas\int \frac{d^2\,x_{02}}{2 \pi} \frac{ x^2_{01}}{x^2_{02}\,x^2_{12}} \Delta^D_{0} \Lb x_{02}\Rb \Delta^D_{0}\Lb x_{12}\Rb -  \Delta^D_{0} \intl^{x^2_{01}} \frac{d x^2_{02}}{x^2_{02}}  \Delta^D_{02}
\eeq
For numerical estimation we have to insert  $ \Delta^D_0$ in the non-linear term and the particular  solution in the region I with geometric scaling  $\Delta^D_1 $  is given by
 \beq \label{2I7} 
 \Delta^D_1( z, z_0)\,\,=\,\,- \Delta^D_0\Lb z, z_0\Rb \intl^\infty_z\,d z' \frac{1}{ \Delta^D_0\Lb z', z_0\Rb}\,\mathscr{N_{L}}[\Delta^D_{0}\Lb z'\Rb]\;.
 \eeq
In   \fig{m} we present the numerical estimates for       
$\mathscr{N_{L}}$  and  the ratio  $\Delta^D_1 /\Delta^D_0 $ turns out to be small. 
           In the region  II     $\Delta^D_1\Lb z, \xi, z_0\Rb$    has been found in our previous paper \cite{CLMNEW} 
(see Eq. 58-67)  and   $\mathscr{N_{L}}[\Delta^D] $ has the same form as  \eq{2I1} with $\Delta^{(0,II)}_0$.
  Finally, the solution $\Delta^D_1\Lb z, \xi, z_0\Rb$    has the following form:
    \bea \label{RIID18}  
   \Delta^D_1\Lb z, \xi, z_0\Rb  \sim \Delta^D_0\Lb z, \xi, z_0\Rb  \left(-\text{erf}\Lb\frac{z_0 - 4\ln2}{\sqrt{2 \kappa }}\Rb+\text{erf} \Lb \frac{z- 4\ln2}{\sqrt{2 \kappa}  } \Rb\Rb\;,
      \eea
    which    vanishes at $z \to z_0$ and in the region of large $z$ this contribution  is rather small.

\section{Conclusions}

Using the modified  homotopy approach we    found that   the first iteration of the homotopy  approach  gives the main contribution in  both kinematic regions which we consider   for the diffractive production. We also   found that for $\xi \,<\,\xi^A_0$   our solution shows the geometric scaling behaviour,  while for $\xi \,>\,\xi^A_0$   this behaviour is strongly violated.   We found that the analytical solution of the non-linear equation, reproduces the initial and boundary conditions. The second iteration with zero initial and boundary conditions turns out to be small and could be taken into account together with higher iterations using the regular perturbative procedure.

\section*{Acknowledgments}
 This research was supported by Fondecyt (Chile) grant No. 1231829.
J.G.  express his gratitude to the PhD scholarship USM-DP No. 029/2024.

\bibliographystyle{unsrt}


\end{document}